\documentstyle[11pt,aaspp4,flushrt]{article} 
\input psfig
\def\simg{\mathrel{\hbox{\rlap{\lower.55ex \hbox {$\sim$}}
                   \kern-.3em \raise.4ex \hbox{$>$}}}}
\def\siml{\mathrel{\hbox{\rlap{\lower.55ex \hbox {$\sim$}}
                   \kern-.3em \raise.4ex \hbox{$<$}}}}
\def\Mesz{M\'esz\'aros~}

\begin{document}
\title{Photon acceleration in variable ultra-relativistic outflows and \\
 high-energy spectra of Gamma-Ray Bursts.} 
\author{Andrei Gruzinov $^1$ and Peter \Mesz $^{1,2}$}
\affil{$^1$ Institute for Advanced Study, School of Natural Sciences, Princeton, 
NJ 08540}
\affil{$^2$ Pennsylvania State University,  525 Davey Lab., University Park, PA 16802~~~~~~~}

\begin{center} DRAFT:~~ \today \end{center}

\begin{abstract}
MeV seed photons produced in shocks in a variable ultra-relativistic outflow 
gain energy by the Fermi mechanism, because the photons Compton scatter off 
relativistically colliding shells. The Fermi-modified high-energy photon spectrum has a 
non-universal slope and a universal cutoff. A significant increase in the total radiative
efficiency is possible.  In some gamma ray bursts, most of the power might be emitted 
at the high-energy cutoff for this mechanism, which would be close to 
100 MeV for outflows with a mean bulk Lorentz factor of 100.
\end{abstract}

\keywords{Gamma-rays: Bursts --- Radiation Mechanisms}

\section{Introduction}

The most common model of gamma-ray burst (GRB) sources involves a relativistic outflow
in which shocks occur and radiate away a fraction of the bulk kinetic energy.  
For typical model parameters the synchrotron emission peaks around 1 keV in the 
comoving frame, or $\sim 100$ keV in the observer frame.  Generally, this is considered 
to be the primary spectrum observed.  However, in internal shocks in the neighborhood 
of the flow photosphere, and also in external shocks in some cases (e.g. Madau, Blandford,
\& Rees 2000), the shocks can have a non-negligible Thomson scattering depth, which 
results in upscattering of these primary photons. Single scattering on
individual shock-accelerated electrons with Lorentz factors $\gamma _e\sim 300$ produces 
photons with energies $\sim \gamma _e^2$ keV in the comoving, or $\sim \Gamma \gamma _e^2
\sim 10\Gamma _2$ GeV in the observer frame, where $\Gamma =100\Gamma _2$ is the bulk 
Lorentz factor.  

Here we concentrate on a different, multiple scattering component. This is associated with 
mildly relativistic motions of different ejecta shells or turbulent cells resulting from 
multiple shock interactions, which contain more energy than the shock-accelerated highly 
relativistic electrons. Multiple interacting shells are naturally expected in internal shocks, 
and also in external shocks when a longer lasting modulated outflow runs into the first 
decelerated shell (e.g. Fenimore \& Ramirez-Ruiz 2000; Kumar \& Piran 2000).

Repeated scatterings using the energy of these bulk motions boost the photon energy through 
the equivalent of the Fermi acceleration mechanism of particles (Blandford \& Payne 1981). 
This is related to Thompson's (1994) photon scattering off Alvf\'en waves, but our mechanism 
relies instead on relative bulk motions. It differs also in using synchrotron photons 
instead of thermal photons as its source term, and hence leads to different characteristic 
energies. This bulk Comptonization results in a spectrum extending at least up to $\sim$ MeV 
in the comoving frame and $\sim 100\Gamma_2$ MeV in the observer frame. The spectral power 
or luminosity per decade can increase as steeply as linearly in the photon energy. This 
provides a natural explanation for those GRB spectra (e.g. Preece et al 1999) which show a 
positive $\nu F_\nu$ slope above the MeV range (generally $\leq +1$), which cannot be 
explained by direct synchrotron radiation from Fermi shock-accelerated electrons. 

The component made up of bulk-scattered photons can extend up to a maximum observed energy 
$\sim 100 \Gamma_2$ MeV. Beyond this energy, Klein-Nishina and electron recoil effects 
set in, and the spectrum reverts to being dominated by the seed spectrum (with negative or
flat power law slope) of the unscattered photons above the synchrotron peak. 

In this work we adopt a test photon model, which assumes no back reaction on the plasma. 
Several potentially important effects are neglected (upscattered photons can heat electrons in 
the colliding shells and produce pairs, light pressure ensures that the total comoving energy 
of the scattered photons cannot exceed the total kinetic energy of the relative shell 
motions). These effects will be investigated elswehere (Gruzinov, \Mesz, \& Rees 2000).  
The simplified approach that we use here retains the essential properties of the bulk motion 
comptonization phenomenon, and allows us to explore the main qualitative features it
introduces in the spectra.

In \S 2 we specify the GRB model, in \S 3 we give an analytical model of Fermi acceleration 
of photons by colliding shell, and in \S 4 we describe our Monte Carlo simulations. 
The results are discussed and related to current and future observations in \S 5.

\section{The GRB internal shock model}

As a specific example to illustrate the effect, we restrict ourselves here to the internal
shock model. We use a standard set of parameters for the ultra-relativistic outflow model 
of GRBs (e.g. \Mesz \& Rees 2000), i.e. luminosity $L=10^{52}L_{52}$ erg/s, $L_{52}\sim 1$; 
terminal Lorentz factor $\Gamma \sim 100$; variability time scale $t_v=10^{-3}t_{ms}$ s,  
with $t_{ms}\sim 1$. With these parameters the shock between shells ejected at typical time 
intervals $t_v$ with $\Delta\Gamma\sim\Gamma$ occurs at a radius $R\sim \Gamma ^2ct_v$. 
To determine the optical depth to Thomson scattering, we estimate the proper density of the 
wind as 
\begin{equation}
n\sim {L\over 4\pi R^2\Gamma ^2m_pc^3}.
\end{equation}
The proper radial width of colliding shells is $\sim R/\Gamma$, and the optical depth is 
\begin{equation}
\tau ~\sim ~n\sigma _TR/\Gamma ~\sim ~{L_{52}\over t_{ms}}\left( {200\over 
\Gamma }\right) ^5.
\end{equation}
We see that optical depths of $\simg 0.1$ are natural. In fact, values of  $\tau \sim 1$ 
are not purely coincidental, since in the standard GRB model the parameters are chosen so 
as to make $\tau <1$ enabling a non-thermal spectrum, and the ``preferred'' model uses 
smaller values of $\Gamma$ (which do not require overly low baryon loads), thus making 
$\tau \siml 1$. The observational fits to the models provide reasonable support for such a 
choice of parameters (e.g. van Paradijs, Kouveliotou \& Wijers, 2000).

Most of the seed photons are emitted at the synchrotron peak. Assuming mildly relativistic internal shocks, and magnetic 
field and electron energies equal to a fraction $\xi _B$ and $\xi _e$ of their 
equipartition values, the synchrotron peak frequency in the comoving frame is 
\begin{equation}
h \nu ~\sim ~\left( {\xi _e\over 0.1}\right) ^2\left( {\xi _B\over 0.1}\right) ^{1/2}
 {L_{52}^{1/2}\over t_{ms}}\left( {200\over \Gamma }\right)^3{\rm keV}.
\end{equation}

\section{Analytical model of photon acceleration by colliding shells}

The basic features of the bulk Comptonization process can be understood by means of a
simple analytical model, which reproduces the essence of the Monte Carlo simulation
discussed in \S 4. This model is sufficient to show that the high-energy slope of 
$\nu F_{\nu }$ can be positive, and gives a value for the cut off energy. 

\begin{figure}[htb]
\psfig{figure=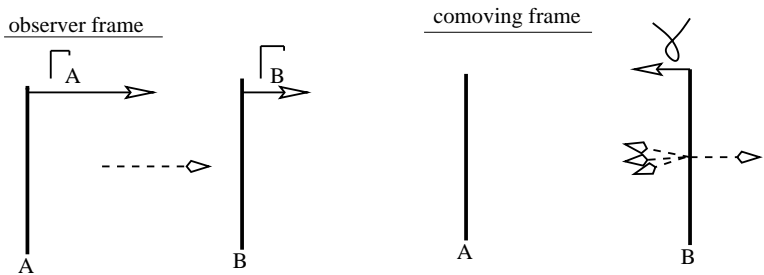,width=4in}
\caption{A photon scatters off relativistically colliding shells. As seen by shell A, 
the scattered photon has a small angle of incidence.}
\label{fig:1}
\end{figure}

Consider a photon between two shells of equal optical depth $\tau \ll 1$ 
(Figure \ref{fig:1}). The shells collide with an ultra-relativistic relative Lorentz 
factor $\gamma \gg 1$ 
\footnote{$\gamma ={1\over 2}\left( {\Gamma _A\over \Gamma _B} + {\Gamma _B\over 
\Gamma _A}\right) $. The case of $\gamma \sim 1$ requires a more cumbersome 
analysis. At our accuracy, we just assume that the ultra-relativistic solution 
is a useful approximation in the mildly relativistic regime.}. Initially, in the frame 
comoving with shell B, the photon frequency is $\nu$ and the incidence angle is 0. 
We assume the Thomson regime, $h \nu \ll m_ec^2$.  The probability that the photon 
passes through shell B without scattering is $1-\tau$. The photon scatters forward with 
probability $\tau /2$, and it scatters backward with probability $\tau /2$. In the frame 
comoving with shell A, the backward scattered photon has a small incidence angle. The 
average (over the scattering kernel) of the Lorentz-transformed frequency is
\begin{equation}
\langle \nu \rangle ~=~ 
   {\int _{\pi /2}^{\pi } d\theta \sin \theta (1+\cos ^2\theta )(1-\cos \theta )
                  \over \int _{\pi /2}^{\pi } d\theta \sin \theta (1+\cos ^2\theta )} ~\gamma \nu~=~
 {25\over 16}~\gamma \nu~.
\label{eq:omav}
\end{equation}

Thus, after one scattering the total energy of photons is changed by a factor of 
$(25/32)\tau \gamma$, and the frequency is changed by a factor of $(25/16)\gamma$. Then the 
observed spectrum is a power law with the luminosity per frequency octave 
$\nu F_{\nu }\propto \nu ^{\beta }$, with
\begin{equation}
\beta ={\log (25\tau \gamma /32)\over \log (25\gamma /16)}~.
\label{eq:beta}
\end{equation}
The slope is positive if $\tau \gamma >1.28$. The steepest slope is $\beta =1$, which is 
achieved in the limit $\gamma \gg \tau ^{-1}$. This model is similar to that for the 
ultrarelativistic isotropic multiple scattering (c.f. Rybicki \& Lightman 1979, p. 212) 
where the slope is $\beta=1+\log (\tau ') /\log(\gamma '^2)$. In the last equation we 
should use $\tau '\sim \tau$ and $\gamma '=\sqrt{(1+\gamma )/2}$, which is the Lorentz 
factor of the colliding shells in the center of mass frame. 

A finite convergence time of the two geometrically thin shells does not introduce a limit on the maximum 
photon energy, as can be seen from the equivalent of Zeno's paradox (since the photons 
always travel faster than the shells approach, the number of photon reflections can in 
principle grow arbitrarily large for a shrinking separation). However, the effective number 
of scatterings should be limited, introducing a corresponding spectral cut off at high 
energies, due to (i) a gradual Klein-Nishina decline 
of the cross section; (ii) a sharp cut-off of the energy of scattered photons due to 
electron recoil (the maximal energy of a photon backward scattered from shell B observed 
in frame A is $\gamma m_ec^2$);  (iii) radiation pressure  back reaction effects 
(if $\beta >0$, the total energy of the scattered photons builds up and the 
photon pressure can prevent the collision of the shells). 

\begin{figure}[htb]
\psfig{figure=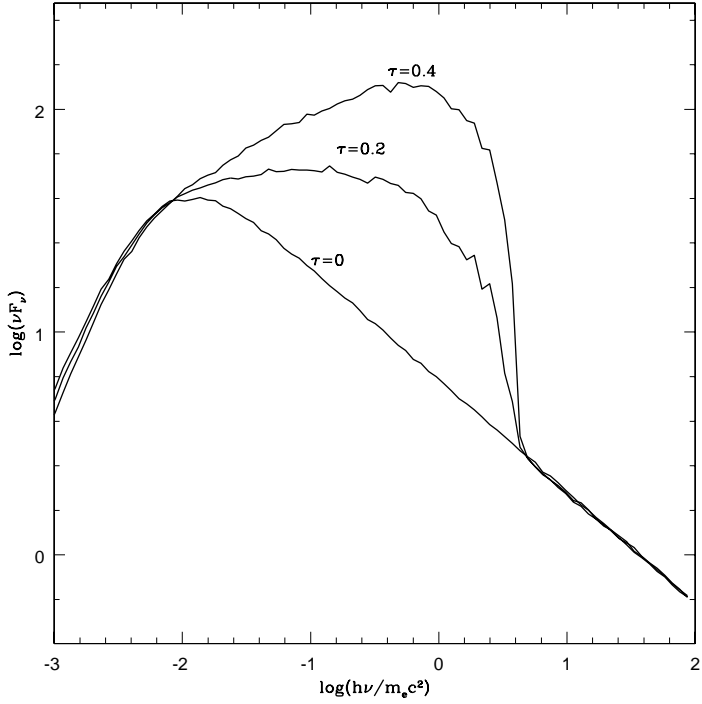,width=4in}
\caption{Monte Carlo simulations of bulk Comptonization between two converging slabs 
(relative Lorentz factor $\gamma =5$), showing the comoving flux per logarithmic photon 
energy interval for three different optical depths $\tau$. The observed spectrum is the
same multiplied by the mean bulk Lorentz factor $\Gamma=10^2\Gamma_2$.}
\label{fig:2}
\end{figure}

\section{Monte Carlo simulations}

Monte Carlo simulations of the photon acceleration model sketched in Figure \ref{fig:1}
were performed in order to check the simple estimates of \S 3. These confirm that the
analytical results, including equation (\ref{eq:beta}), are a useful approximation in 
the mildly relativistic regime. They also allow us to find in more detail the shape of 
the Compton recoil / Klein-Nishina cut off (see Figure \ref{fig:2}). The Monte Carlo 
simulations were done using two approximations, which simplify the problem without affecting 
significantly the result: polarization effects are neglected, and 
multiple scattering during one passage through a shell is neglected. 

Initially, a seed synchrotron spectrum is released. The seed spectrum is isotropic in 
the center of mass frame, rising at low energies ($\beta =4/3$), with a smooth transition
to a spectrum decreasing at high energies (where we take as an example $\beta =-0.5$). 

In Figure \ref{fig:2} we show a spectral calculation for two shells similar to those
of Figure {\ref{fig:1}, using a relative Lorentz factor $\gamma=5$. The unscattered seed 
synchrotron spectrum is marked ``$\tau =0$''. The spectrum of all photons
that finally escape outwards (to the right) through the shell B,  as measured in the
(comoving) frame B, is shown in Figure \ref{fig:2} for different values of $\tau$.
The Compton recoil / Klein-Nishina effects are noticeable at about 1 MeV 
comoving energies, and there is a sharp Compton recoil cut off at $h\nu /(m_ec^2)=
\gamma $. In the observer frame this spectrum would be blueshifted by the factor $\Gamma_B$.

\begin{figure}[htb]
\psfig{figure=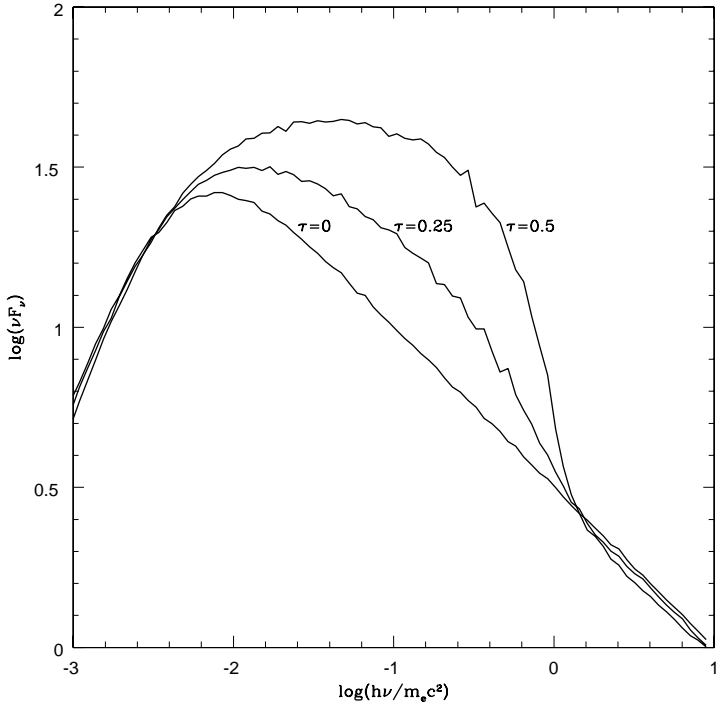,width=4in}
\caption{Same as figure \ref{fig:2}, comoving frame spectrum for a case with $\gamma =2$.}
\label{fig:3}
\end{figure}

The dependence on the relative Lorentz factor is illustrated in Figure \ref{fig:3},
showing the effect for a fiducial relative Lorentz factor $\gamma =2$, which may also be 
characteristic of the reverse shock in an external deceleration shock scenario.
As expected, the steepening remains present but becomes less strong as weaker shocks
are considered. The cut-off, however, is inherent to the scattering physics, and 
although it has a tendency to decrease somewhat with decreasing $\gamma$, its order of 
magnitude remains in the neighborhood of $h\nu \sim m_e c^2 \sim 1$ MeV (comoving frame).

\section{Discussion}

The Fermi acceleration of photons in the standard GRB fireball model has several
consequences of theoretical and observational interest. The first is that it can 
naturally produce $\nu F_\nu$ spectra which are harder than the input synchrotron spectrum,
including the possibility of an increasing $\nu F_{\nu }$ above the usual break found 
in the Band parameterization of spectra (Band et al 1999, Preece et al 1999). The latter
is a property that is hard to obtain with a synchrotron model. Such rising $\nu F_\nu \propto
\nu^\beta$ are expected, in this model, to have $\beta \siml 1$, and this implies that
in some GRB most of the energy is at energies $h\nu \sim 100 (\Gamma /10^2)$ MeV, 
well above the BATSE instrument band on the Compton Gamma Ray Observatory (CGRO). The current 
observational situation is that BATSE finds approximately 16 \% of the spectra to be rising 
at $\simg 1$ MeV, and the observed rise is not faster than  $\beta =1$ (Preece et al 1999). 
These fits generally cut off above 1.8 MeV, and the break is usually not much lower than 
this, so there is some uncertainty. The COMPTEL instrument on CGRO, sensitive up to 30 MeV,
has analyzed $\sim 30$ bursts (Schoenfelder et al 2000), and an analysis of the slopes 
indicates in several cases $\nu F_\nu$ slopes $\beta\sim 0$, with one burst of $\beta\sim 
0.5$ (Kippen et al 1999). The EGRET experiment on CGRO has detected $\sim 30$ bursts with 
the scintillation counters in the 1-200 MeV range, and $\sim$ 7 bursts with the spark 
chambers in the 100 MeV-30 GeV range. In this range the spectra are largely noise-dominated 
(Schaefer et al 1998, Bromm \& Schaefer 1999). However, the scintillation spectral slopes 
(Catelli, Dingus \& Schneid, 1997) are compatible with $\beta\siml 0$, although some could 
be positive and others negative, which is compatible with the analysis of spark chamber data
(e.g. Sommers, 1994; Hurley et al, 1994). Large area detectors such as GLAST should be able 
to obtain more definite answers in the 20 MeV-300 GeV range.

Another implication of the photon acceleration described here is that it provides a natural 
mechanism to increase the efficiency of conversion of baryon bulk motion into photon energy. 
This is of interest since in general the internal shock synchrotron efficiency for radiating 
in the BATSE band is limited to 1-10\% (Kumar 2000; Spada, Panaitescu \& \Mesz 2000; 
see however Fenimore \& Ramirez-Ruiz 2000, and observation-based estimates by Freedman \& 
Waxman 2000). 
The increase in the radiative efficiency is simply given by the increase in the value of
$\nu F_\nu$ at different energies, or by the integral $\int F_\nu d\nu$ in the range of
interest. In the generic examples  shown, this increase is substantial.  
Of course, the spectra shown in Figures \ref{fig:2} and \ref{fig:3} are test photon spectra, 
which do not take into account the back-reaction of radiation. The latter can become important 
when a substantial fraction of the bulk energy has been converted to radiation through
Fermi acceleration, and the natural limit for this effect can be estimated as $\sim 50\%$
radiative efficiency. 

The calculations presented here are meant to illustrate the consequences of photon 
acceleration by bulk motions. A self-consistent calculation is needed in order 
to explore the back-reaction of the radiation pressure and pair formation on the
shell dynamics.  Pair formation has an angle averaged cross section 
$\sigma _{\gamma \gamma}\sim \sigma_T/8$ at a threshold which is similar to that 
for Klein-Nishina effects. It is not very important in low compactness situations (shocks 
at large radii), while in high compactness cases the exponential tail of the photon 
distribution would lead to a pair cascade, which lowers the cut off in the comoving spectra 
of Figures \ref{fig:2},\ref{fig:3} to $\siml$ 0.5 MeV, possibly with some pile-up of 
photons at this energy (Gruzinov, \Mesz \& Rees, 2000).

\acknowledgements{We thank V. Connaughton, B. Dingus, P. Kumar and M.J. Rees for 
valuable comments. AG was supported by the W. M. Keck Foundation and NSF PHY-9513835. 
PM was supported by NASA NAG5-2857, the Guggenheim Foundation and the Institute for 
Advanced Study.}

\end{document}